\documentclass[sigconf]{acmart}

\makeatletter
\renewcommand\@formatdoi[1]{\ignorespaces}
\makeatother
\renewcommand\footnotetextcopyrightpermission[1]{} 
\usepackage{booktabs} 

\newtheorem{remark}{Remark} 

\usepackage{amsmath}
\usepackage{algorithm}
\usepackage[noend]{algpseudocode}
\usepackage{graphicx}

\usepackage{subfloat}
\usepackage{subfig}
\usepackage[utf8]{inputenc}
\usepackage[T1]{fontenc}
\usepackage{lmodern}
\usepackage{amsthm}
\usepackage{tabularx}

\usepackage{enumitem}

\newcolumntype{L}{>{\raggedright}X}

\begin{document}
\title[A Fast Fragmentation Algorithm]{A Fast Fragmentation Algorithm For Data Protection In a Multi-Cloud Environment}

\setcopyright{none}

\author{Katarzyna Kapusta and Gerard Memmi}
\affiliation{\institution{LTCI, Telecom ParisTech, Universite Paris-Saclay}}
\email{katarzyna.kapusta,gerard.memmi@telecom-paristech.fr}


\begin{abstract}

Data fragmentation and dispersal over multiple clouds is a way of data protection against \textit{honest-but-curious} storage or service providers. In this paper, we introduce a novel algorithm for data fragmentation that is particularly well adapted to be used in a multi-cloud environment. An empirical security analysis was performed on data sets provided by a large enterprise and shows that the scheme achieves good data protection. A performance comparison with published related works demonstrates it can be more than twice faster than the fastest of the relevant fragmentation techniques, while producing reasonable storage overhead.


\end{abstract}

%
%



\maketitle

\section{Introduction}
\label{sec: intoduction}

Data protection inside cloud-based storage systems is quite a challenge. Data breaches happen on a daily basis, in extreme cases leading to massive disclosure of sensitive data \footnote{https://www.bloomberg.com/news/features/2017-09-29/the-equifax-hack-has-all-the-hallmarks-of-state-sponsored-pros}. A user uploading his data to the cloud has to trust that the provider will be able to protect the data not only from exterior attackers, but also from insider threats like malicious system administrators or implementation issues. Moreover, a user will have to accept the fact that in some situations the cloud provider will not disclose neither in which country his data have been stored nor if this data have been revealed to a foreign administration or third party. Encrypting data before  uploading them to the cloud seems to be a solution to the problem. However, it requires that the user will take care of the key management. A different approach consist in fragmenting data and dispersing them over multiple clouds in a way that no single cloud get enough information to recover the data.

Data protection by means of fragmentation is not a new idea. Since the early 90' it appears in various proposals of architectures for distributed storage \cite{notere}. The development of the cloud technology reinforced the interest in this methodology, as it allows a broad dispersal of data over multiple servers on different sites. Several authors already proposed to use not one but two or more independent storage providers to increase protection or availability of the data. Some of the solutions fragments data along their structure in order to break the dependencies \cite{aggarwal,hudic}, while the rest apply different fragmentation techniques from data shredding to perfect or computational secret sharing \cite{depsky,cdstore,systemx,future:internet}. The choice of the right fragmentation technique depends on the use case, as it is usually a compromise between the desired level of data protection, the produced storage overhead, and the fragmentation performance. In this paper we introduce a novel fragmentation algorithm adapted for the use in a multi-cloud environment. It fills the need for a fast keyless fragmentation method, that would provide a good level of protection without a drastic increase in storage.

We consider a user who wants to keep his data confidential, but he wants the solution to be fast, scalable, and inexpensive. To comply with this last requirement, he will at least partially outsource his data to a public environment such as a cloud. Then, one of his problems regarding privacy becomes to efficiently protect his data from the curiosity of the provider. We propose to address this issue by fragmenting the user’s data, encoding them, and dispersing fragments over several non-colluding sites. 

A new scheme situated between computational secret sharing and information dispersal algorithms was sketched out in our short three pages poster~\cite{Kapusta:2016}. It has a limited domain of application, because of its lack of scalability (performance strongly depends on the desired number of fragments) and a level of data protection that could be easily improved. In this paper, we present a thoroughly revisited version of this scheme that deals with scalability and security issues. 

This paper is organized as follows. After rapidly describing our motivations and our contribution in Section~\ref{sec:contribution}, as well as relevant works in Section \ref{sec:relevant},  we detail our algorithm and its data model (Section \ref{sec:concepts} and \ref{sec:algorithm-description}).Then we analyze it in terms of security (Section \ref{sec:theoretical-security-evaluation} and Section \ref{sec:empirical-security-evaluation}),  as well as complexity and memory occupation (Section \ref{sec:theoretical-performance}). Last but no least, we benchmark it against the state-of-the-art techniques in Section \ref{sec:measured-performance}. An insight into future works ends up the paper.

\section{Contribution and motivations}
\label{sec:contribution}

During the description of the proposed fragmentation method, as well as during the presentation of the relevant works, the following notations will be used:
\begin{itemize}[leftmargin=*]
\item[-] $d$: initial data of size $|d|$
\item[-] $f$: a data fragment of size $|f|$ coming from the fragmentation of $d$
\item[-] $c$: the number of separated locations i.e. non-colluding cloud providers
\item[-] $k$: the minimum number of fragments required for data recovery; $k$ is known in the literature as a threshold  
\item[-] $n$: the total number of fragments (including fragments added for availability)
\end{itemize}

The proposed algorithm fragments initial data $d$ into $k$ fragments $f_0,...,f_{k-1}$ that will be dispersed and stored over $c$ non-colluding clouds. Data recovery is therefore not possible unless all $k$ fragments are gathered. More precisely, the fragmentation process uses a combination of secret sharing and data permuting to create dependencies between data inside the fragments. Such $k$ fragments are then dispersed over $c$ clouds. The dispersal follows special rules ensuring that the fragments stored at one provider do not reveal the content of the initial data $d$. Additional $n-k$ fragments could be generated. However, we assume that in most cases there is no need for that, as the cloud providers already apply their own methods, i.e. replication, to ensure data availability.
The proposed fragmentation method achieves better performance than the state-of-the art techniques, while generating a reasonable memory overhead, and providing a good level of data protection.

A typical use case for the application of the scheme is the situation when a user would like to upload his data to a public cloud storage service, but does not completely trust the cloud provider. Therefore, he prefers to disperse his data between two or three clouds in a way that any of cloud does not get enough information about the data to recover them. 

In the proposed threat model, we consider that a single cloud provider is \textit{honest\--but\--curious}: he will try to look at the data he was entrusted with, but will not make the effort to contact the $c-1$ other cloud providers (who are supposed to be unknown to him) in an attempt to recover the data. A cloud site may also be vulnerable to exterior attacks leading to a data leakage. In such a situation, the goal of the algorithm is to fragment the data between the clouds in a way that the fragments received by a single cloud are practically useless.  

Another assumption in our use case, is that in a situation where the choice of an appropriate fragmentation method is a compromise between performance, memory overhead, and data protection the user favors the performance. Indeed, if the data are very sensitive or even critical, additional protection method could be applied, like perfect secret sharing or symmetric encryption (obviously this would decrease the performance of the solution).  In such a case, some fragments could be stored in a private and trusted site, leading to an increasing of storage costs.

\section{Related works to data fragmentation}
\label{sec:relevant}

We selected four relevant works from the domain since they are widely used by both researchers and engineers. Two of them are precise algorithm descriptions (Shamir's secret sharing and Information Dispersal Algorithm), while the other two (Secret Sharing Made Short and AONT-RS) are rather flexible fragmentation methods where one cryptographic mechanism can be replaced by another one. Later in this paper, we will present comparisons of these algorithms together with our own proposition in terms of security, as well as in term of memory occupation and performance. A more complete state of the art can be found in \cite{notere} showing that these techniques had a vast impact and have been successfully transfered in the industry.

\subsection{Shamir's secret sharing (SSS)}
\label{ssec:shamir}

Shamir's secret sharing scheme ~\cite{shamir:1979} takes as input data $d$ of size $|d|$ and fragments them into $n$ fragments $f_0,...,f_{n-1}$ of size $|d|$ each, that will be later distributed by a dealer to $n$ different players. Any $k$ of these fragments are needed for the recovery of $d$. The algorithm is based on the fact that given $k$ unequal points $x_0,...,x_{k-1}$ and $k$ arbitrary values $y_0,...,y_{k-1}$ there is at most one polynomial $p$ of degree less or equal to $k-1$ such that $p\left(x_i\right)=y_i, i=0,...,{k-1}$. In SSS data are transformed into $n$ points $f_0=\left(x_0,y_0\right),...,f_{n-1}=\left(x_{n-1},y_{n-1}\right)$ belonging to a random polynomial. The scheme provides information-theoretic security, but has quadratic complexity in function of $k$ and leads to a n-fold increase of storage space. It is usually applied for protection of small or critical data like encryption keys. In such a use case, drawbacks of the scheme are acceptable or negligible, but for larger data they become a major stumbling block. In \cite{parakh:space-efficient}, a space efficient secret sharing scheme was presented that modifies the Shamir's scheme in order to save memory occupation. Shamir's scheme does not protect against a malicious dealer that would like to modify the data fragments. A verifiable secret sharing scheme in which the players can verify if their fragments are consistent, was introduced in \cite{vss}.

\subsection{Information Dispersal Algorithm (IDA)}
\label{ssec:ida}  

Rabin's Information Dispersal Algorithm~\cite{rabin:1989} divides data $d$ of size $|d|$ into $n$ fragments of size $\frac{|d|}{k}$ each, so that any $k$ fragments suffice for reconstruction. More precisely, the $n$ data fragments are obtained by multiplying the initial data by a $k \times n$ nonsingular generator matrix. The $n$ rows of the matrix are dispersed within the data fragments. Recovery consists in multiplying any $k$ fragments by the inverse of a $k \times k$ matrix built from $k$ rows of the generator matrix received within the $k$ fragments. An IDA adds redundancy to data and produces a negligible storage overhead. It provides some data protection, as the input data cannot be explicitly reconstructed from fewer than the $k$ required fragments~\cite{Li:ida}. However, even if it is not possible to directly recover the data, some information about the content of the initial data is leaked. Indeed, data patterns are preserved inside the fragments when the same matrix is reused to encode different data. A similar problem occurs while using the Electronic Code Book block cipher mode for block cipher symmetric encryption~\cite{Dworkin:2001:SER:2206247}. Even with this weakness, it is still being considered as one the techniques that could be potentially useful in a multi-cloud environment~\cite{future:internet}. 


\subsection{Secret Sharing Made Short (SSMS)}
\label{subs:ssms}

Krawczyk's Secret Sharing Made Short ~\cite{krawczyk:1993} combines symmetric encryption with perfect secret sharing and introduces a secret sharing adapted for protection of larger data. In this method, data $d$ are first encrypted using a symmetric encryption algorithm, then fragmented using an Information Dispersal Algorithm. The encryption key is fragmented using a perfect secret sharing scheme (like SSS) and dispersed within data fragments. In consequence, the solution does not require explicit key management. The storage overhead does not depend on data size $|d|$, but is equal to the size of the key per data fragment. The performance of the SSMS technique depends on the details of the chosen techniques for encryption and data dispersal.

\subsection{AONT-RS}
\label{ssec:aont-rs}

Similarly to SSMS, the AONT-RS method~\cite{Resch:2011,revisiting-aont-rs} combines symmetric encryption with data dispersal. The difference between the two methods lies in the key management. Inside AONT-RS data $d$ are first encrypted using a symmetric key algorithm and then the key used for encryption is XOR-ed with the hash of the encrypted data. Such processed data are then fragmented into $k$ fragments and $n-k$ fragments are added using a systematic Reed-Solomon error correction code \cite{RS} (which may be seen as an IDA). The performance of the AONT-RS technique depends on the details of the chosen techniques for encryption, data hashing, and the Reed-Solomon implementation. A CAONT-RS method is similar to the AONT-RS, except that the key used to encrypt the data is not random, but generated from the hash of the input data \cite{cdstore,convergent-dispersal}.

The AONT-RS took inspiration from the all-or-nothing transform (AONT) introduced by Rivest ~\cite{rivest:97}. An AONT is a preprocessing step applied before encryption, that makes impossible to recover the input data unless possessing the entire ciphertext. Input data are seen as a sequence of messages, that are encrypted with a random key and then the key is XOR-ed with the hashes of the messages. The cost of an AONT proposed by Rivest is approximately twice the cost of the actual encryption, in contrary to AONT-RS where the encryption is applied only once. 




\section{Data concepts, definitions, and prerequisites}
\label{sec:concepts}

This Section introduces few data concepts and definitions that may help the understanding of the description of the fragmentation algorithm. We also point out few prerequisites necessary for an optimal execution of the fragmentation.

\subsection{Data concepts}

Our approach utilizes the following key data components with their size in number of bits and their dimensions in terms of number of blocks (and for a block or a share in terms of mini-blocks and mini-shares respectively):

\begin{itemize}[leftmargin=*]
\item Mini-block ($mb$): a sequence of bits of size $|mb|$
\item Data block or block ($b$): a sequence of bits of size $|b|$ belonging to the original input data and for ease of computation will be a multiple of the size of a mini-block, therefore, containing $\#b=\frac{|b|}{|mb|}$ mini-blocks 
\item Data ($d$): an input data of size $|d|$ bits composed of $\#d=\frac{|d|}{|b|}$  data blocks; assuming that the size of $d$ is a multiple of the size of a block (if ever that was not to be the case, some padding technique could be applied)
\item Mini-share ($ms$): a sequence of bits of size $|ms|=|mb|$ generated by the transformation of a mini-block
\item Share ($s$): a sequence of bits of size $|s|=|b|$ bits, coming from the transformation of a data block and containing $\#s=\#b$ mini-shares 
\item Permutation array ($pa$): an array of size $|pa|$ bits containing $\#pa=\#b$ values: all natural numbers in range $[0,...,\#b-1]$ appearing in a random order
\item Permutation share ($ps$): a share of size $|ps|=|pa|$ bits coming from a splitting of a permutation array $pa$
\item Fragment ($f$): a fragment composed of $\#f=\frac{|f|}{|b|}=\frac{\#d}{k}$ blocks at the beginning of the algorithm; then composed of the same number $\#f$  of data shares plus one permutation share at the end of the algorithm. 
\end{itemize}
In the sequel, we will assume that $d$ and its size and $c$ the number of independent storage sites are given by the use case description and that the user will be choosing the threshold $k$ as a multiple of $c$ and that at the same time in order to simplify the reading of this paper, the size of $d$ can be considered as a multiple of the size of a mini-block verifying the following equation: $|d|=k\#f\#b|mb|$.

\subsubsection{Notations} 
\label{ssec:notations}

A fragment $j$ is denoted by $f_j$. A block $i$ inside a fragment $f_j$ is denoted by $b^j_i$.
A share $i$ inside a fragment $f_j$ generated through the transformation of the block $b^j_i$  is denoted by $s^j_i$.  A permutation array $pa_r$ is used to permute mini-shares inside the fragment $f_j$, where $r$ is an integer in $[0,...,\frac{k}{c}-1]$ such that $j \pmod{\frac{k}{c}}=r$.
A value at the position $t$ inside a permutation array $pa_r$ is denoted by $pa_r\left(t\right)$.
A permutation share $z$ coming from the split of a permutation array $pa_r$ into $c$ permutation shares is denoted by $ps_{r,z}$, where $z$ is an integer in $[0,...,c-1]$ such that $j\pmod{c}=z$.  
A mini-block at the position $v$ inside a block $b^j_i$ is denoted by $b^j_i\left(v\right)$, where $v$ is an integer in $[0,...,\#b-1]$. 
A mini-share generated from a mini-block $b^j_i\left(v\right)$, at the position $w$ inside a share $s^j_i$ is denoted by $s^j_i\left(w\right)$, where $w$ is an integer in $[0,...,\#b-1]$ such that $pa_{rz}\left(v\right)=w$.

\subsection{Definitions}
\label{sec:definitions}

The core of our fragmentation process consists in encoding a mini-block using already encoded mini-blocks (mini-shares), constituting a new mini-share.  Mini-share after mini-share, shares are then constituted. Doing so is creating dependencies between data at the level of mini-shares, of shares, and ultimately of fragments. Such created dependencies are the strongest only between groups of $c$ fragments that will be called neighbors. On the one hand, these dependencies are assuring that $k$ fragments are necessary to reconstruct meaningful data. On another hand, they can be felt as a key information to undertake cryptanalysis. To minimize the opportunity for an attacker to use them, neighbor fragments will be systematically dispersed to different independent sites. Shares and mini-shares contained in neighbor fragments and used to generate a new share or mini-share from a block or a mini-block will be called parent share or mini-share along the definitions hereunder. All $k$ fragments are therefore constructed share by share, row of shares after row of share.

By convention, a permutation share $ps_{r,z}$ will be considered as a share $s^j_0$ with $j=rc+z$. A  share $s^j_i$  is the result of encoding of the block $b^j_i$ using $c-1$ previously encoded shares from $c-1$ other fragments. The $c-1$ shares used for the encoding of  $s^j_i$ are its parent shares and the fragments to which they  belong are the neighbor fragments of the fragment $f_j$. Similarly, $c-1$ mini-shares used to encode a mini-share  $s^j_i\left(w\right)$  are called parent mini-shares and they belong to parent shares of  $s^j_i$ contained inside the neighbor fragments of  $f_j$.  Intuitively, dependencies between a share/mini-share and its parents or between a fragment and its neighbor are stronger than between the rest of the data. These dependencies could be exploited by an attacker and have to be taken care of. Later in this section, we describe how these relations will be broken by dispersing concerned data over independent sites.

\begin{definition}{\textbf{Parent shares}}
A share $s^j_i$ belonging to a fragment $f_j$ such that $i>0$ possesses $c-1$ parent shares: \newline ${s^{(j+1)\bmod k}_{i-1},..,s^{(j+c-1)\bmod k}_{i-1}}$
\label{def:parent-shares} 
\end{definition}

\begin{definition}{\textbf{Parent mini-shares}}
A mini-share $s^j_i\left(w\right)$ belonging to the share $s^j_i$, that was generated from the mini-block at the position $v$ inside the block $b^j_i$ and such that $i>0$, possesses $c-1$ parent mini-shares inside the parent shares of $s^j_i$:
\newline
  ${s^{(j+1)\mod k}_{i-1}\left(v\right),..,s^{(j+c-1)\mod k}_{i-1}\left(v\right)}$
\label{def:parent-ms} 
\end{definition}

\begin{definition}{\textbf{Neighbor fragments}}
A fragment $f_j$ from the set of $k$ fragments $f_0,...,f_{k-1}$ possesses $c-1$ neighbor fragments: \newline
${f_{(j+1)\bmod{k}},..,f_{(j+c-1)\bmod k}}$ 
\label{def:neighbor-fragment}
\end{definition}

\subsection{Prerequisites}
\label{ssec:prerequisites}

\textbf{Size of data $d$} Data $d$ of size $|d|$ are composed of $\#d$ blocks of data ${b_1,...,b_{\#d}}$. Therefore, the size of the data is equal to $|d|=|b|\times\#d$. During fragmentation, the data are distributed over $k$ fragments in a way that each of the fragments receives a portion of data of size $\frac{|d|}{k}$. To ensure a balanced distribution, the number of blocks inside the data $\#d$ has to be a multiple of the number of fragments $k$. This can be achieved by the use of padding.

\textbf{Number of mini-block inside a block} The chosen value $\#b$ of number of mini-blocks inside a block has an impact on the size of the fragments. A permutation array has to contain all the natural values from the range of  $[0,...,\#b-1]$. If the number of mini-blocks inside a block $\#b$ is greater than the maximum value that can be encoded on $|mb|$ bits, the size of permutation arrays is greater than the size of the blocks. To simplify computations and minimize memory overhead, the maximum size of the block should not greater than the maximum value that can be represented on $|mb|$ bits. 

\textbf{Values of k and c} $k$ defines the number of fragments required for data recovery and $c$ represents the number of cloud providers. Obviously, $k$ have to be greater than $c$. In order to facilitate the computations, $k$ should be also a multiple of $c$. This comes from the fact that during the fragmentation process $k$ shares are produced from $c$ permutation arrays that have to be later dispersed over $c$ clouds.

\section{Creating and dispersing fragments}
\label{sec:algorithm-description}


Following subsections detail how our fragmentation algorithm transforms data $d$ into $k$ fragments and how it disperses the $k$ fragments over $c$ clouds. 
The pseudo-code of the whole procedure can be found in Figure~\ref{fig:transformingDataIntoFragments}. We do not describe the defragmentation procedure as it is an inverse of the fragmentation.

\subsection{Data distribution over fragments}
\label{ssec:dataDispersal}

\begin{figure}[!ht]
\centering
\includegraphics[width=0.98\linewidth]{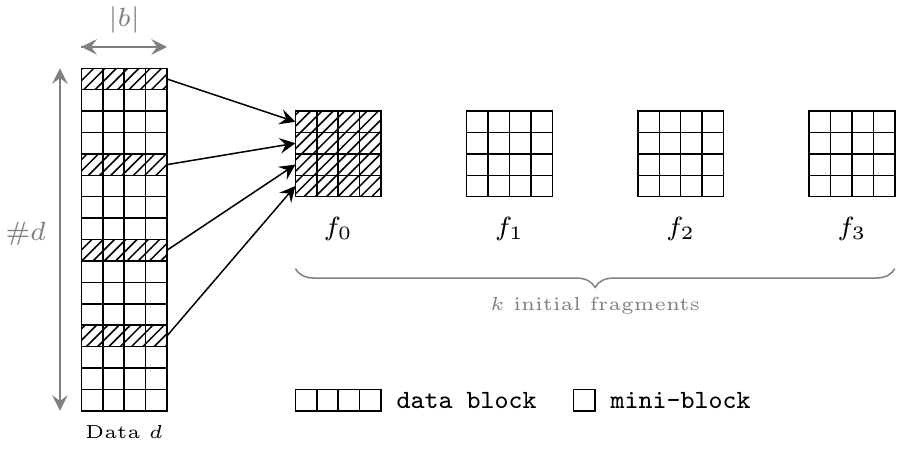}  
\caption{{\it Distributing data blocks over $k=4$ initial data fragments. Each fragment receives $\frac{1}{4}$ of the data $d$. Any pair of adjacent blocks are distributed in different fragments.}}%
\label{fig:initial}%
\end{figure}

In a first step implemented inside the \textsc{FormFragments} function, data $d$ are distributed over $k$ fragments $f_0,...,f_{k-1}$ in such a way that  $b_i$ is assigned to $f_j  \Longleftrightarrow i \bmod k = j$. Because the number of blocks is a multiple of $k$, each fragment receives exactly $\frac{|d|}{k}$ of the input data (see schema Figure \ref{fig:initial}). This method of proceeding was chosen, as it allows to start the encoding of first distributed data blocks before the whole data are distributed over fragments in a pipelined manner. The distribution of data over fragments could also be performed in a simpler way: data could be just divided into $k$ consecutive chunks of size $\frac{|d|}{k}$. 

\subsection{Generating permutations}
\label{ssec:permutations}

Permutation arrays are used during data encoding, therefore they have to be generated before the start of the encoding procedure. \textsc{GeneratePermutations} function generates $\frac{k}{c}$ random permutation arrays $pa_0,...,pa_{\frac{k}{c}-1}$ of length $\#pa=\#b$ containing all natural numbers from the range $[0,...,\#pa-1]$ appearing in a random order.  

The main role of a permutation array is to mix up positions of mini-shares inside a freshly generated share. The encoding creates dependencies between mini-shares from neighbors fragments, so it is not possible to recover a mini-share inside a fragment without possessing its parents mini-shares from neighbor fragments. However, the proposed encoding is not a complex operation, as its main goal is to be fast. An attacker may try to guess the values of $c-1$ missing mini-shares in order to recover a single mini-block. The fact that he knows that the order of the encoded mini-shares is the same as the order of mini-blocks inside initial data may be helpful in this attack attempt. Changing positions of mini-shares inside a share slows down an attacker, as in addition to guessing values from fragments that he does not possess he has to obtain the right order of the encoded data. With a reasonably long permutation array it becomes a significant obstacle on the way to partial data recovery.

\begin{figure}[h]
\centering
\includegraphics[width=0.98\linewidth]{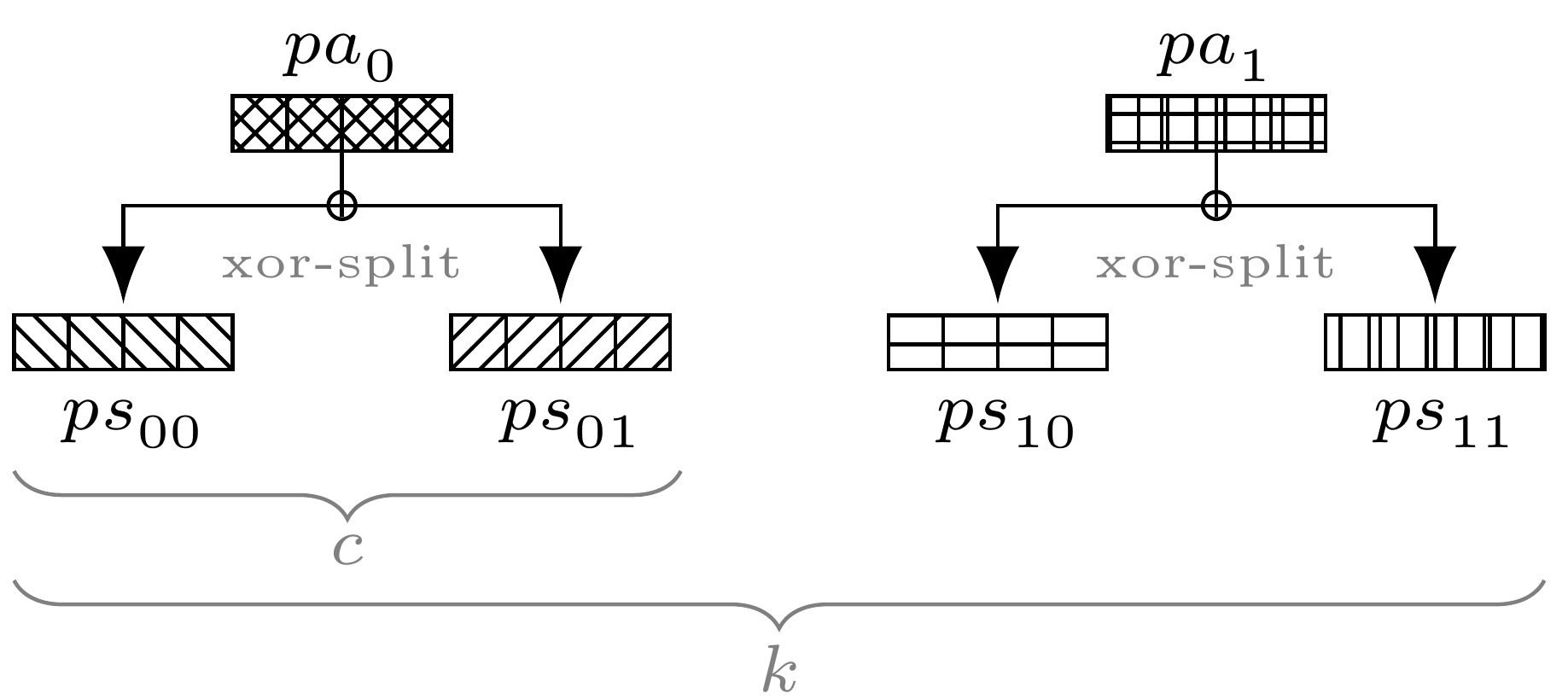}  
\caption{{\it Splitting $\frac{k}{c}$ permutation arrays into $k$ permutation shares. Here, $c=2$ and $k=4$. As an example, the permutation share $ps_{1,0}$ will be appended to the fragment $f_2$.}} 
\label{fig:permutations}%
\end{figure}

Permutation arrays are necessary for data recovery and important for data protection, therefore they have to be securely transmitted within the final $k$ data fragments. 
In the proposed algorithm, they are fragmented using a perfect secret sharing scheme into $c$ permutation shares each that will be dispersed over $c$ cloud providers.
In more details, the function \textsc{Split\-Permutations} implements a XOR-split that splits permutation arrays into $c$ permutation shares each (presented in Figure \ref{fig:permutations}). The obtained $k$ (because $\frac{k}{c}\times c=k$) permutation shares are then distributed over $k$ fragments in a way that $ps_{rz}$ is assigned to $f_j  \Longleftrightarrow r\times c + z = j$.  A permutation share $ps_{rz}$ becomes the first share $s^j_0$ of a fragment $j$. At the end, $k$ fragments containing in total $\frac{k}{c}$ sets of permutations shares will be dispersed over $c$  clouds. This observation leads to the formulation of the Remark \ref{remark:permutation} on the secure dispersal of the fragments.

\begin{remark}[Dispersing permutation shares]
\label{remark:permutation}
Fragments that store one of the $c$ permutation shares belonging to one single permutation array should be stored separately, i.e. dispersed over different $c$ cloud providers.
\end{remark}

Each permutation share plays a double role. It is not only used to recover one of the permutation arrays, but as it is the first share of a fragment, it is also used as a kind of initialization block during data encoding.


\subsection{Data encoding}
\label{ssec:dataEncoding}

\begin{figure}[!ht]
\centering
\includegraphics[width=0.85\linewidth]{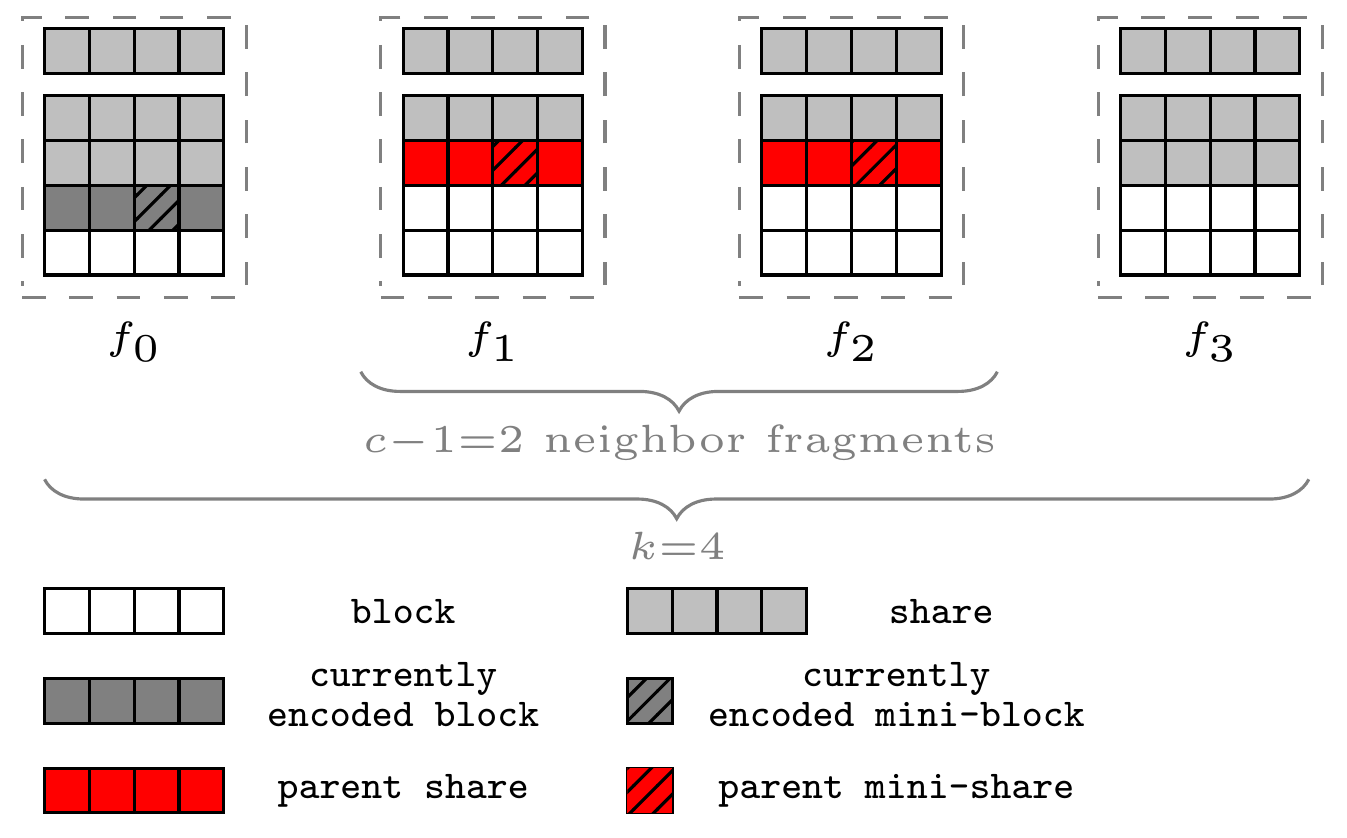}  
\caption{{\it Encoding data blocks into data shares.  }}%
\label{fig:encodingBlocks}%
\end{figure}

After data dispersal described in Section \ref{ssec:dataDispersal}, each fragment contains a part of the input data that could potentially reveal some information about the initial data content. Therefore, data inside fragments have to be encoded. The goal of the encoding is to create relationships between neighbor fragments at the level of shares and mini-shares, in a way that it is not possible to recover any block or any mini-block of initial data without gathering all of the $k$ fragments. To achieve this, each block $b^j_i$ inside an fragment $f_j$ is encoded using $c-1$ parent shares from neighbor fragments of $f_j$ (illustrated in Figure \ref{fig:encodingBlocks}). The processing is sequential and its philosophy could be roughly compared to the CBC encryption mode: once a block is encoded into a share it becomes a parent to the next block. The first block of each of the fragments does not posses natural parents, as its the first block to be encoded into a share (a similar problem occurs in CBC mode, where an initialization vector is introduced as the first block). For the $k$ first blocks permutation shares are used as parents. This solution has two advantages. First, it allows to save storage space and not to generate $k$ fresh initialization shares. Second, because permutation shares are pseudo-random (they are the result of a xor-split) they add some randomness to the encoding. The processing is sequential and requires that all the $k$ fragments are encoded at the same time. An encoding of a block into a share consists in encoding each of its mini-block using mini-shares from $c-1$ parent shares. After a mini-share is encoded, its position inside its block is permuted with the help of one of the permutation arrays. A block with all its mini-shares encoded becomes a share.

\subsubsection{Encoding a block into a share}

A block $b^j_i$ is transformed into a corresponding share $s^j_i$ inside the function \textsc{Encode\-Block} . Each of the mini-block inside a block will be encoded in a Shamir's like fashion as a point of a polynomial of degree $c-1$, where the coefficients of the polynomial are picked from parent shares of the block.

Before the start of the block encoding, parent shares ($ParentShares$) of the block are selected. Following the Definition~\ref{def:parent-shares}, a share $s_{i-1}^l,l={(j+1)\bmod k,\ldots,(j+c-1)\bmod k}$, becomes a parent share of $b^j_i$ (function \textsc{Select\-Parent\-Shares}). Intuitively, parent shares are the last $c-1$ shares from the neighbors of the fragment that is being encoded (or $c-1$ permutation shares when the block to be encoded is the first block of a fragment). Parent shares will be used for selecting coefficients of the encoding polynomials. Moreover, an evaluation point $x$ is selected that is an integer in range of $[2,...,2^{|mb|-1}]$ ($2^{|mb|-1}$ being the maximum value that can be encoded on $|mb|$ bits) (function \textsc{PickX}). $x$ will be the point at which the encoding polynomials for mini-shares will be evaluated. In addition, $Encode\-Block$ takes also as input the permutation array $pa_{j \mod{\frac{k}{c}}}$ ($PermArray$) that will be used for permuting mini-shares.

\begin{figure}[h!]
\begin{algorithmic}[1]
\Function{EncodeBlock}{$ParentShares$,$PermArray$,$x$,$b^j_i$}
    \For {$v = 0:\#b-1$}
	\State $ParentMS$=\textsc{SelectParentMS$\left(ParentShares\right)$}
	\State  $ms$ = \textsc{EncodeMiniBlock$\left(ParentMS,x,b^j_i\left(v\right)\right)$}
    \State  $s^j_i\left(w\right)$ =\textsc{PermuteMiniShare}$\left(PermArray\left(v\right),ms\right)$
\EndFor
\EndFunction
\end{algorithmic}
\caption{Pseudo-code of the function \textsc{EncodeBlock} that transforms a block $b^j_i$ into a share $s^j_i$.}
\label{fig:EncodeBlock}
\end{figure}

\begin{figure}[!ht]
\centering
\includegraphics[width=0.85\linewidth]{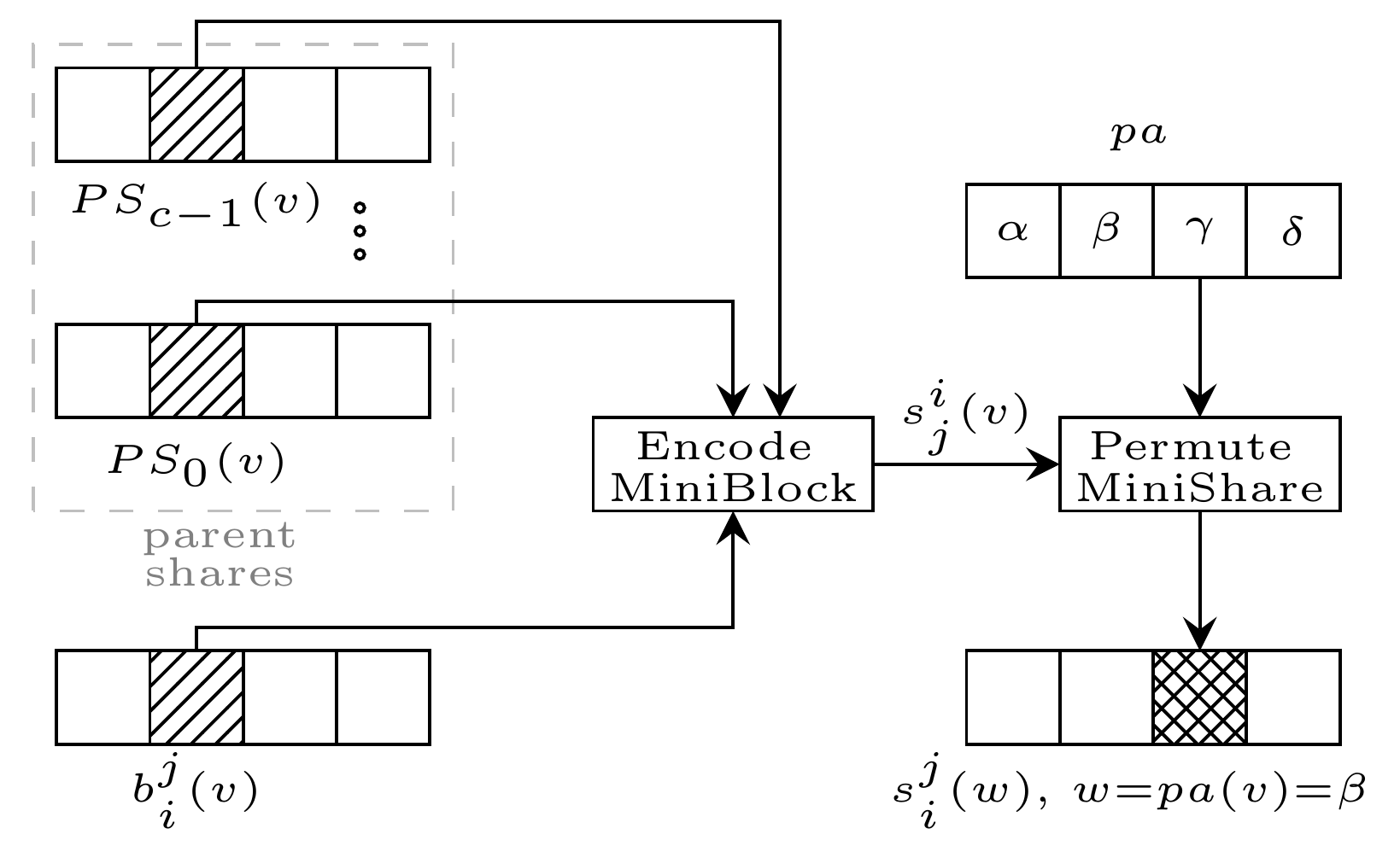}  
\caption{{\it Encoding a mini-block into a mini-share. $ParentsMS$ are used as coefficients of the encoding polynomial, $x$ is the point at which the polynomial is evaluated. }}%
\label{fig:encodeMiniBlock}%
\end{figure}

Once a set of parent shares and the evaluation point are selected, the block is encoded mini-block by mini-block (pseudo-code presented in Figure \ref{fig:EncodeBlock}). For a mini-block $b^j_i\left(v\right)$ a set of $c-1$ parent mini-shares ($ParentMS$) is selected from the set of the parent shares (\textsc{Select\-Parent\-MS}).  Following the Definition~\ref{def:parent-ms}, a mini-share $s^j_{i-1}\left(v\right)$ that is at the position $v$ inside a parent share becomes a parent mini-share. The mini-block $b^j_i\left(v\right)$ is then transformed into a mini-share $s^j_i\left(w\right)$ inside the \textsc{Encode\-Mini\-Block} function. The position $w$ of the mini-share is returned by the function\textsc{Permute\-Mini\-Share}). The process of encoding a mini-block $b^j_i\left(v\right)$ into a mini-share $s^j_i\left(w\right)$ is illustrated in Figure \ref{fig:encodeMiniBlock}.

\subsubsection{Encoding a mini-block into a mini-share}

\textsc{Encode\-Mini\-Block} function (see Figure~\ref{fig:EncodeMiniBlock}) takes as input a mini-block $mb$ to be encoded, $c-1$ parent mini-shares ($ParentMS$), and a value $x$. 

\begin{figure}[h]
\begin{algorithmic}[1]
\State $ParentMS={a_0,a_1,...,a_{c-2}}$
\Function{EncodeMiniBlock}{$ParentMS,x,mb$}
  \State\Return $ms = mb + xa_0 + ... +x^{c-1}a_{c-2}$
\EndFunction
\end{algorithmic}
\caption{Pseudo-code of the \textsc{Encode\-Mini\-Block} function producing a mini-share value $ms$.}
\label{fig:EncodeMiniBlock}
\end{figure}

The encoding procedure is based on Shamir's scheme. A $\left(c,c\right)$ threshold is applied . $c-1$ coefficients are used to construct an encoding polynomial for a single secret (in our case the mini-block to encode constitutes the secret) are not pseudo-random, but they are taken from the set of previously encoded mini-blocks. Such constructed polynomial is evaluated at only one point $x$. Indeed, as the coefficients of the polynomial are also mini-shares, to recover the an encoded mini-block we need its corresponding mini-share and its $c-1$ parent mini-shares used as coefficients. Reusing mini-shares lowers the security level, but prevents an increase in memory (discussed in Section \ref{sec:theoretical-security-evaluation}). Intuitively, a mini-share and its parent mini-shares should not end up inside the same fragments and should be separated from each other. This leads to the following dispersal remark:

\begin{remark}[Dispersing mini-shares]
\label{remark:dispersingminishares}
A mini-share and its $c-1$ parent mini-shares should be dispersed over $c$ different locations (ideally $c$ cloud providers).
\end{remark}

\begin{figure}[h]
\includegraphics[width=0.9\linewidth]{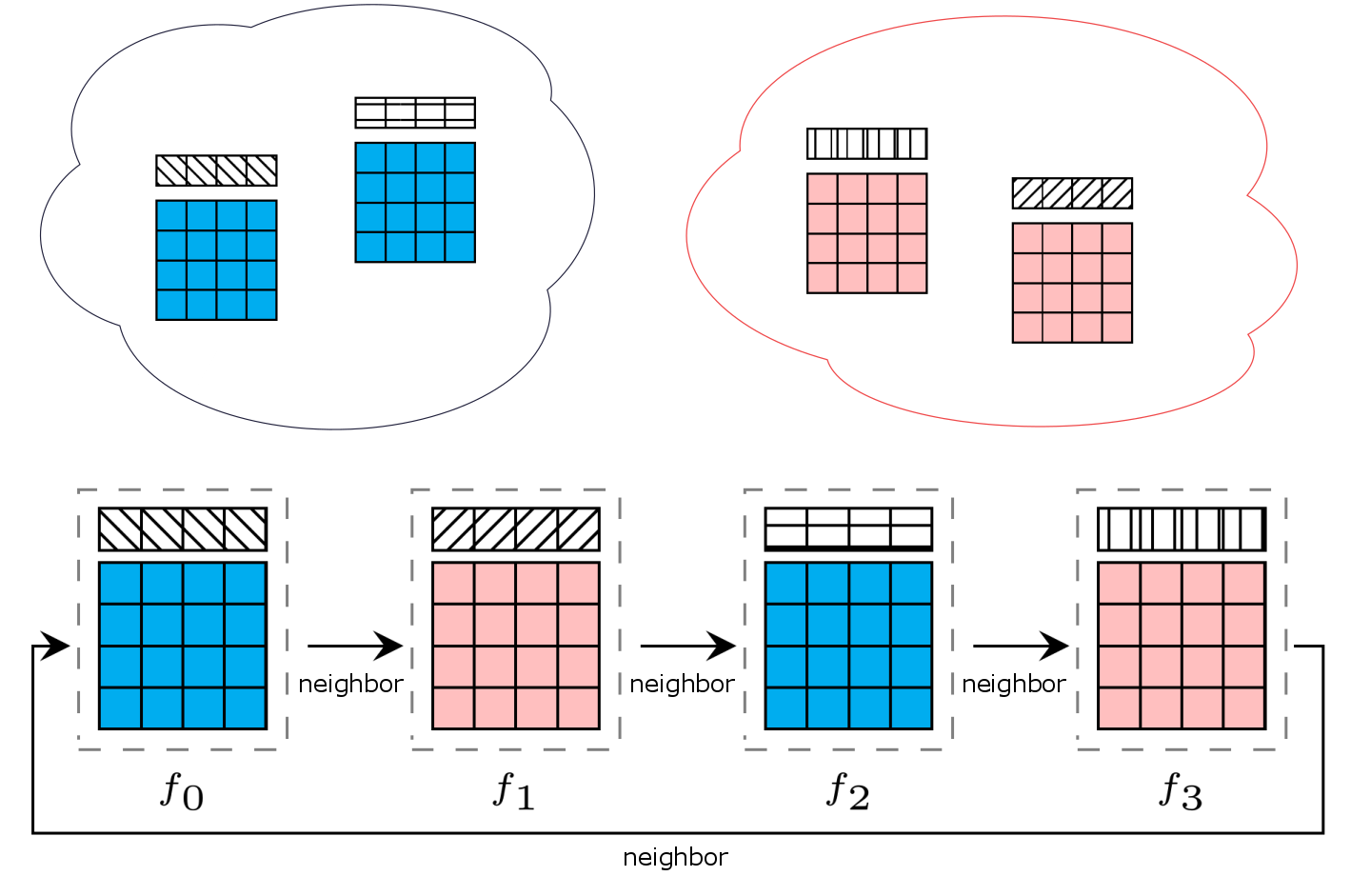}  
\caption{\textit{Fragments dispersal over $c=2$ clouds. Neighbors are dispersed over different clouds.}}
\label{fig:dispersal}
\end{figure}

\begin{figure*}[h]
\begin{algorithmic}[1]
\Function{EncodeData}{$d,c,k$}
\State $f_0,...,f_{k-1}$=\textsc{FormFragments}$\left(k,d\right)$  \Comment{distributes data over $k$ fragments}
\State $pa_0,...,pa_{\frac{k}{c}-1}=$\textsc{GeneratePermutations}$\left(c,k,\#b\right)$  \Comment{ generates $\frac{k}{c}$ permutation arrays}
\State $s^0_0,\ldots,s^{k-1}_0$=\textsc{SplitPermutations}$\left(pa_0,...,pa_{\frac{k}{c}-1}\right)$ \Comment{splits and distributes permutation arrays}
\While {all $\frac{\#b }{k}$ blocks of all $k$ fragments are processed} \Comment{transforms data by sets of $k$ blocks}
\For {each block $b^j_i$ of a fragment $f_j$}
    \State$ParentShares=$\textsc{SelectParentShares}$\left(b^j_i\right)$ \Comment{selects $c-1$ previously generated shares}
	\State $x$=\textsc{PickX}  
    \State $s^j_i$=\textsc{EncodeBlock}$\left(ParentShares,pa_{j\mod{\frac{k}{c}}},x,b^j_i\right)$ \Comment{transforms a block into a share}
\EndFor
\EndWhile
\EndFunction
\end{algorithmic}
\caption{Pseudo-code of the function $EncodeData$ that transform input data $d$ into a set of $k$ fragments.}
\label{fig:transformingDataIntoFragments}
\end{figure*}

\subsection{Dispersing fragments over the clouds}
\label{sec:dispersal}

Data encoding produces $k$ final fragments $f_0,...,f_{k-1}$. A final fragment contains encoded data, as well as one permutation share. Such final fragments are dispersed over $c$ clouds $cloud_0,...,cloud_{c-1}$ (illustrated in Figure \ref{fig:dispersal}). The dispersal procedure (pseudo-code in Figure \ref{fig:dispersal-code}) is defined by one rule: two neighbor fragments cannot be stored at one single cloud provider. The consequence of this rule is the requirement for at least $c$ independent cloud providers. We recommend a $c=2$ or $c=3$, as a larger number of clouds could lead to latency issues. The total number of fragments $k$ is the choice of the user: a higher value of $k$ could reinforce the data protection inside the cloud against external attackers (that would have to gather data fragments not only from different clouds, but also multiple cloud servers). A weaker variation of the dispersal algorithm where only one cloud is used could be also considered: to mislead the cloud a user can upload the data fragments from $c$ different accounts. 
In the considered scenario, we assume that the user does not have to care about data availability, as data availability is usually guaranteed while signing the SLA. A user fearing data loss can however generate $n-k$  additional fragments and disperse them over an additional cloud.

\begin{figure}[H]
\begin{algorithmic}[1]
\Function{DisperseFragments}{$f_0,...,f_j,...,f_{k-1}$}
\For {$i = 0:k-1$}
\State disperse fragment $f_j$ to the cloud $cloud_{j \bmod c}$
\EndFor
\EndFunction
\end{algorithmic}
\caption{Pseudo-code of the \textsc{DisperseFragments} function.}
\label{fig:dispersal-code}
\end{figure}

\section{Security evaluation}
\label{sec:theoretical-security-evaluation}

Each cloud receives $\frac{k}{c}$ non-neighbor fragments. A single \textit{honest\--but\--curious} cloud provider can try two things: to decode a portion of data from received fragments or to verify if data inside received fragments match some presumed data. He has to overcome a combination of three obstacles to satisfy his curiosity: data dispersal, data permuting, and data encoding.

\subsection{Data dispersal} 

The first and simplest obstacle is data dispersal. A single provider receives $k$ fragments containing only a portion of encoded input data of size $\frac{|d|}{c}$ in total. Even decoded,  information contained inside the fragments is sampled (the result of \textsc{Form\-Fragments}) and incomplete. Moreover, if the cloud does not receive any information about the ordering of the fragments, it has to check  $\left(\frac{k}{c}\right)!$ possibilities of fragments reassembling. 

\subsection{Breaking data permuting}

A single cloud receives data fragments, which shares were permuted using one or more permutation arrays. These permutation arrays were split using a perfect secret sharing scheme ($xor$-split) into $c$ permutation shares and each of the cloud providers received only one of permutation shares from the set corresponding to one permutation array. In consequence, a single cloud cannot recover any permutation array. The difficulty of a brute-force attack on a permutation array will obviously depends on the length of the array. For a permutation array of size $\#b$,  $\#b!$ possible permutations exist. If the blocks are small, a brute-force attack is feasible. However, a $\#b=34$ results in $2.95\times 10^{38}$ permutation array possibilities, which is comparable to the number of tries that are required to perform a brute-force attack on a 128-bit symmetric encryption key ($2^{128}$ gives $3.4\times 10^{38}$ possibilities). An increase of the size of the block slightly affects the storage space, but also improves the performance of the fragmentation process (see Section \ref{sec:measured-performance}).
 
\subsection{Breaking data encoding}

Each mini-block of data inside a fragment is encoded using parent mini-shares from neighbor fragments. The dispersal procedure ensures that neighbor fragments (and therefore all the mini-shares implicated in the encoding process) will never end up being stored at the same storage provider. However, parent mini-shares are not totally random, as they depend on the data already encoded, as well as on $c-1$ first pseudo-random blocks. In order to break the data encoding, a cloud can try to solve a system of polynomial equations: each mini-share may be represented in function of its initial mini-block and its parent mini-shares. Recursively, parent mini-shares may be represented in the same way.

If the cloud has some assumptions about the stored data or it would like to verify if its fragments may belong to some predefined data, then its work will be slightly easier. In the first scenario, the assumptions may add additional equations linking the data, but the fact that the encoded data are permuted makes it hard to exploit those dependencies (unless the permutation are being recovered). In the second case, the cloud will only have to guess the missing part of the $\frac{k}{c}$ pseudo-random first shares. With a reasonably large block size $|b|$ it is a non-negligible impediment. 


\section{Empirical security evaluation}
\label{sec:empirical-security-evaluation}

The proposed scheme aims at providing fast data fragmentation that achieves a good data protection level. An experimental analysis of security characteristics of the scheme based on the methodology presented in~\cite{noura} was performed. Tests were adapted to the fragmenting nature of the scheme. Obtained results show that data inside the fragments achieve a good level of uniformity and independence. In contrary to an IDA, the scheme does not preserve patterns inside data fragments.

\begin{figure}[!ht]
\centering
\captionsetup{belowskip=2pt,aboveskip=4pt}
\subfloat[][{\it Original data}]{\includegraphics[width=0.3\linewidth]{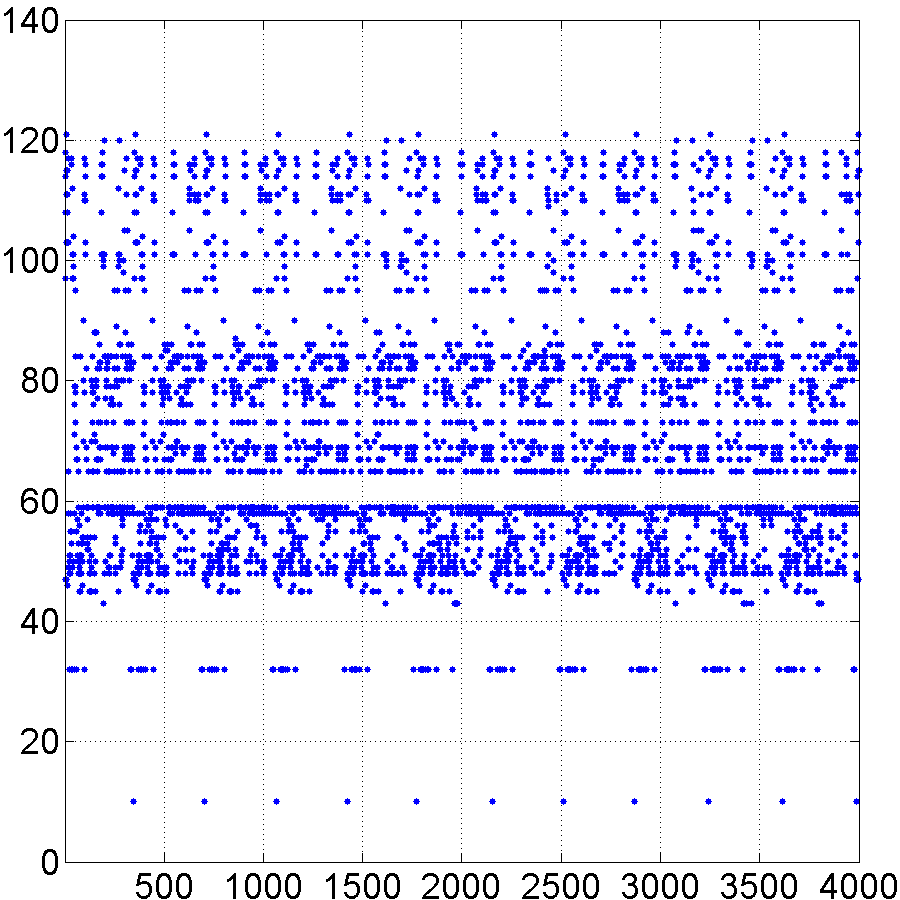}  \label{fig:OM}}  \hfill
\subfloat[][{\it IDA}]{\includegraphics[width=0.3\linewidth]{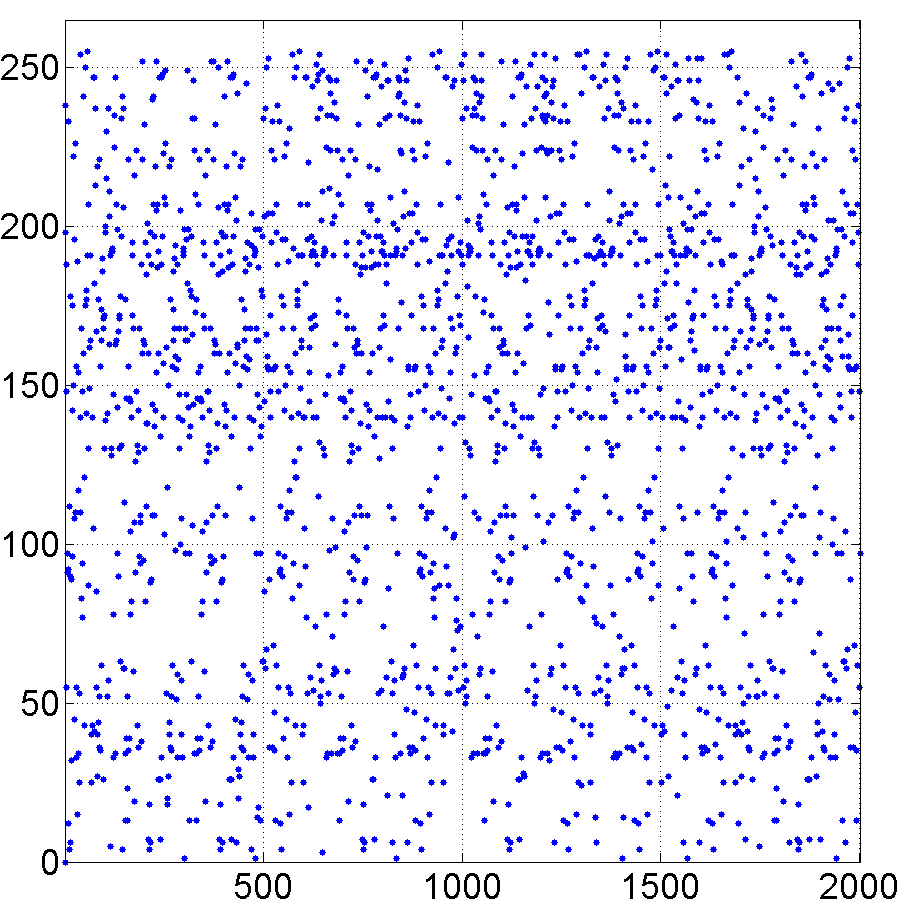}  \label{fig:FRAB}}  \hfill 
\subfloat[][{\it Proposed scheme}]{\includegraphics[width=0.3\linewidth]{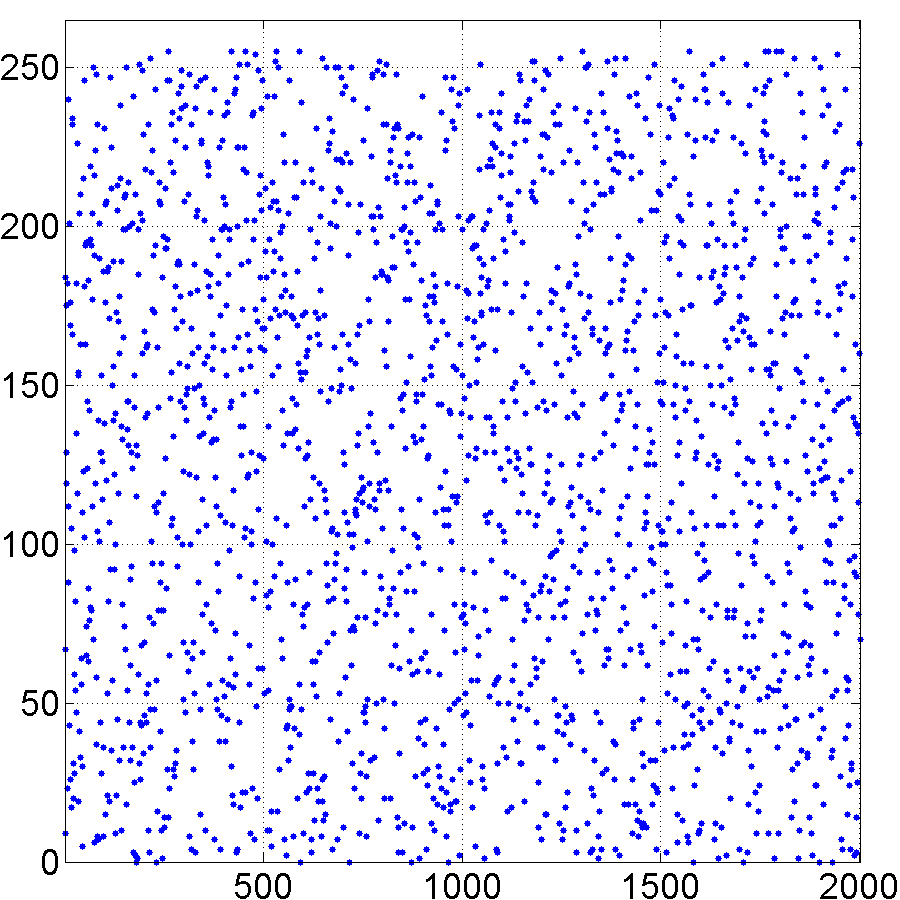} \label{fig:FM}}
\caption{{\it Byte value distribution of a textual data sample (a) and distribution of one of its fragment after applying an IDA (b), and after applying the proposed approach (c), k=2. Data patterns are preserved after the use of an IDA. Fragment (c) contains all possible byte values and does not contain visible data patterns. x-axis shows the byte position inside the sample, y-axis shows the byte value at position $x$.}}%
\label{fig:distr}%
\end{figure}

All tests were performed using Matlab environment on textual data samples provided by the French post office LaPoste\footnote{http://www.laposte.fr/}. An example of one of such data sample is shown in Figure~\ref{fig:OM}. Its corresponding fragment is presented in Figure~\ref{fig:FM} and compared to the one obtained using an IDA (Figure~\ref{fig:FRAB}). 

\begin{figure}[h]
\captionsetup{belowskip=-8pt}
\centering
\includegraphics[width=0.85\linewidth]{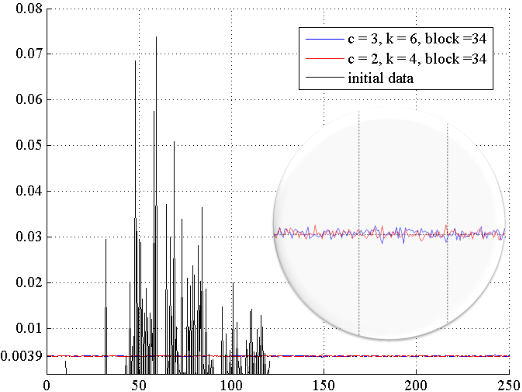} 
\caption{ {\it Probability density function comparison for data and their two different fragments. x-axis shows possible byte values in the sample, the y-axis shows the probability of occurrence of a value. For fragmented data, the occurrence probability is close to the one of a uniform distribution (0.039). } }%
\label{fig:PDF_initial}%
\end{figure}

\noindent \textbf{Probability Density Function.} Frequency counts close to a uniform distribution testify data have a good level of mixing. This means that each byte value inside a fragment should have an occurrence probability close to $\frac{1}{v}=0.0039$, where $v$ is the number of possible values (256 for a byte). In Figure~\ref{fig:PDF_initial}, the probability density function (PDF) of a data sample and two of its fragments are shown. Results for the fragments are spread over the space and have a distribution close to uniform. It demonstrates that the occurrence probability of byte values is close to 0.0039.

\begin{figure}[h]
\centering
\includegraphics[width=0.85\linewidth]{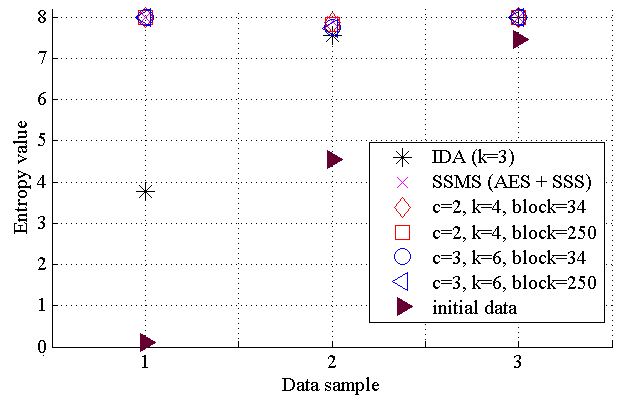}  
\caption{{\it Entropy comparison for three data samples (text, image). Maximum entropy value is equal to 8. Entropy of fragments obtained using IDA depends strongly on the entropy of the input data. Our scheme gives fragments with entropy comparable to other techniques.}}%
\label{fig:entropy}%
\end{figure}

\noindent \textbf{Entropy.} Information entropy is a measure of unpredictability of information content~\cite{entropy}. In a good fragmentation scheme the entropy of the fragments should be close to ideal. Figure \ref{fig:entropy} shows entropy value for three different fragmented data samples for different fragmentation algorithm. The entropy value of the fragments generated using our proposal was comparable with the entropy of fragments generated using SSMS and much higher than the one obtained using an IDA.

\noindent \textbf{Chi-squared test.} Uniformity of byte values of data inside fragments was validated by applying a chi-squared test~\cite{chi2}. For a significance level of 0.05 the null hypothesis is not rejected and the distribution of the fragment data is uniform if $\chi^2_{test} \leq \chi^2_{theory}(255,0.05)\approx293$. The test was applied on fragmentation results of 15 different data samples for a fragment size of 1000 bytes. For all samples the tests was successful.

\begin{figure}[!h]
\centering
\subfloat[][{\it Original data}]{\includegraphics[width=0.3\linewidth]{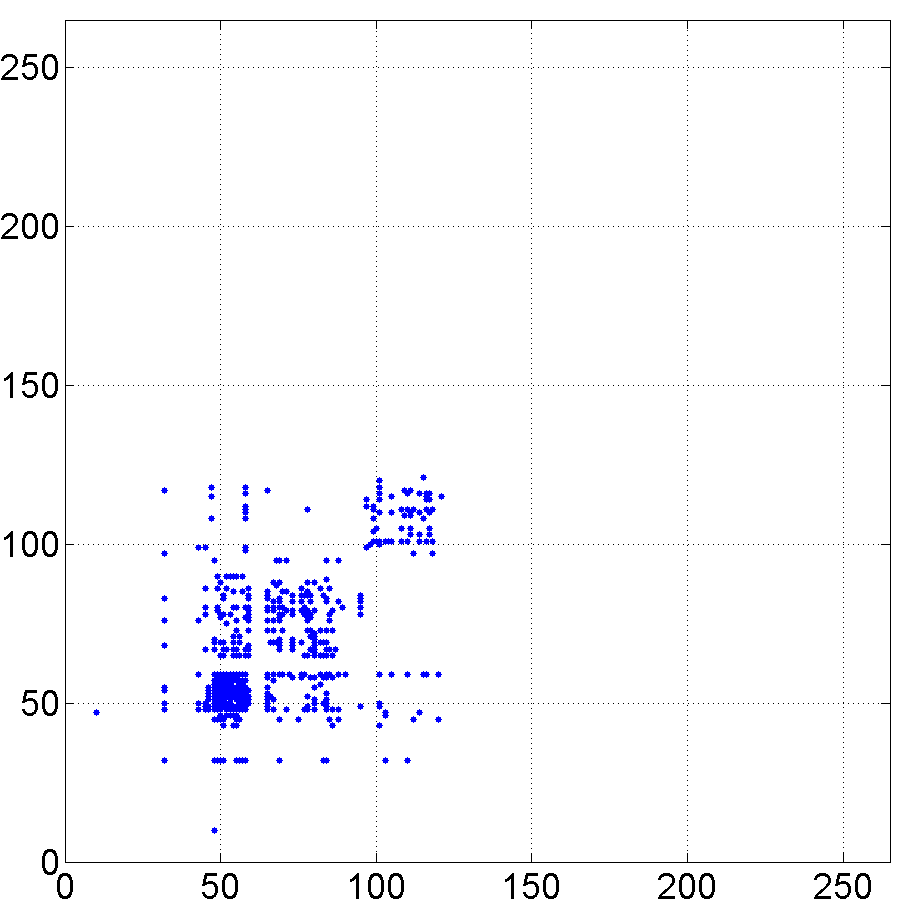}\label{fig:reco}}%
\subfloat[][{\it IDA}]{\includegraphics[width=0.3\linewidth]{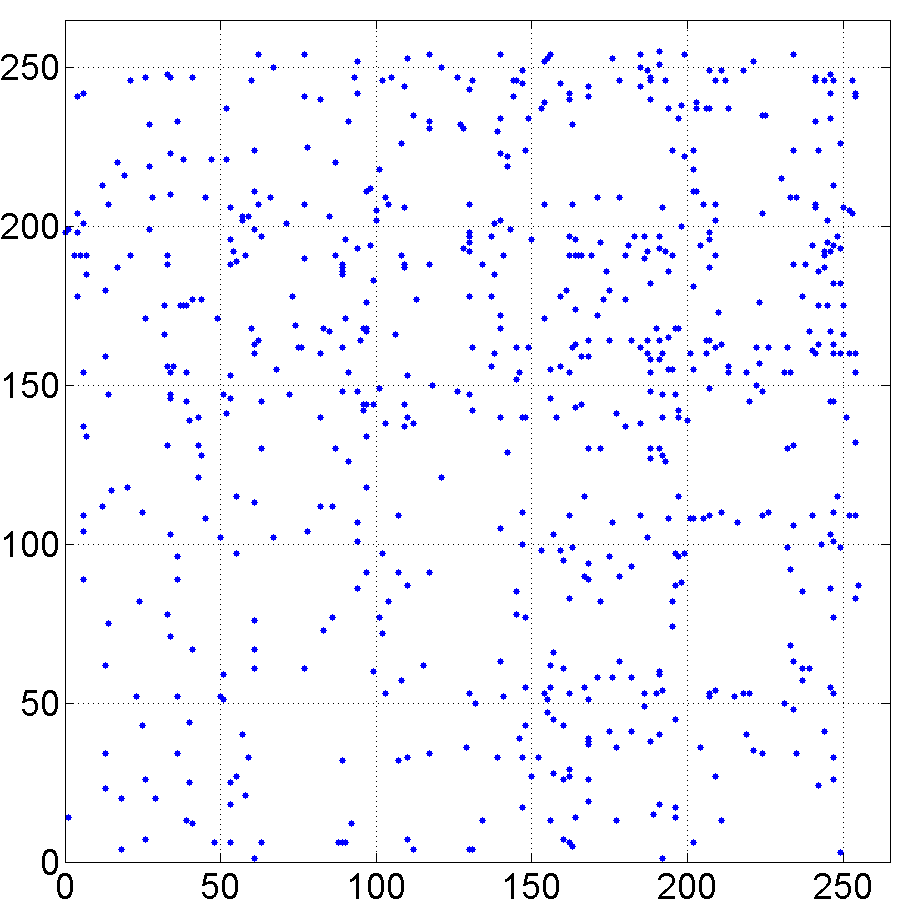}\label{fig:recf}}
\subfloat[][{\it Proposed scheme}]{\includegraphics[width=0.3\linewidth]{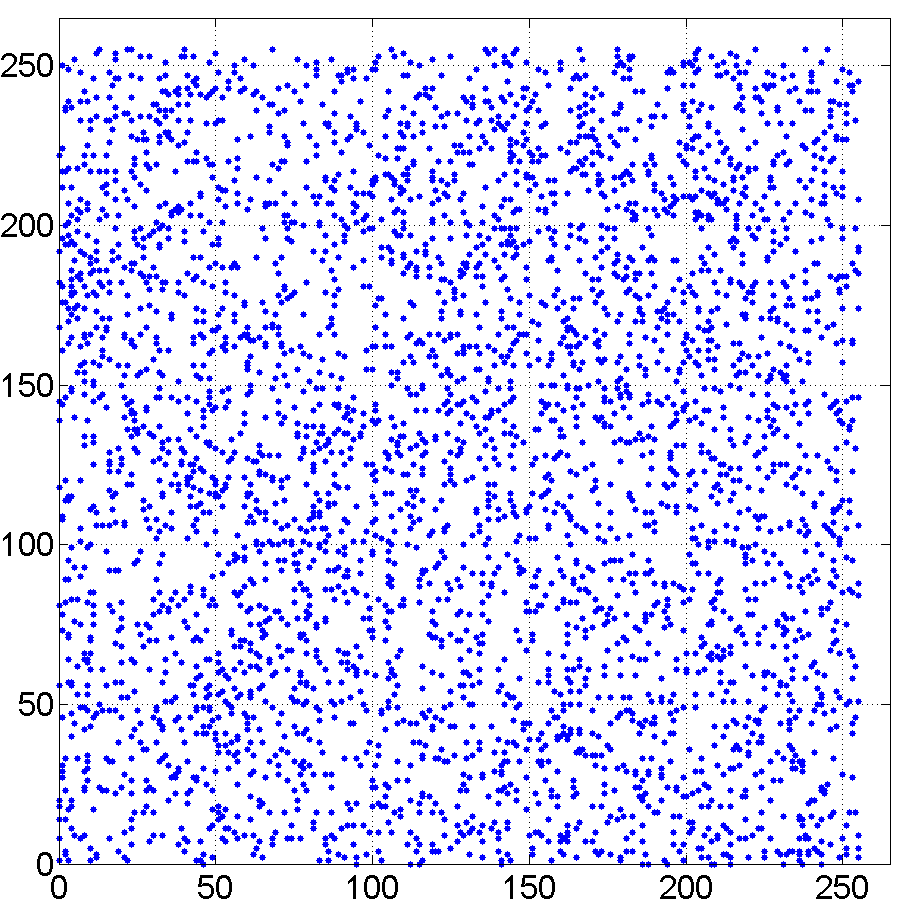}\label{fig:recf}}
\caption{{\it Recurrence plots for data from Figure~\ref{fig:distr}. x-axis shows data values, y-axis shows data values with a delay $t=1$.}}%
\label{fig:REC}%
\end{figure}

\noindent \textbf{Recurrence.} A recurrence plot serves to estimate correlation inside data~\cite{correlation}. Considering data vector $x={x_1,x_2,...,x_m}$ a vector with delay $t \geq 1$ is constructed $x(t)=x_{1+t},x_{2+t},...,x_{m+t}$. A recurrence plot shows the variation between $x$ and $x(t)$. In Figure~\ref{fig:REC}, such plots for a data sample and its fragments obtained by applying an IDA and the proposed scheme are shown. Using the proposed scheme, data inside the fragments are more uniformly distributed.

\begin{figure}[!h]
\centering
\subfloat[][]{\includegraphics[width=0.45\linewidth]{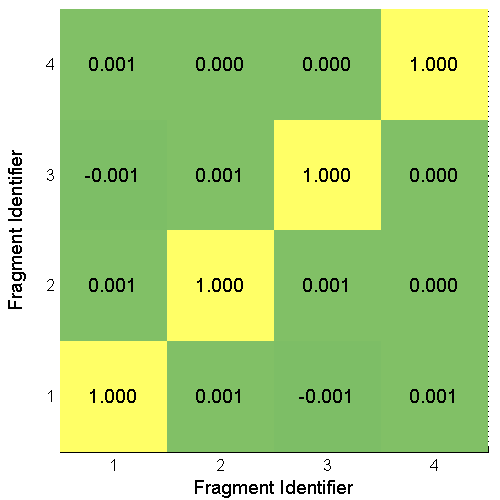} \label{fig:pdf}}
\subfloat[][]{\includegraphics[width=0.45\linewidth]{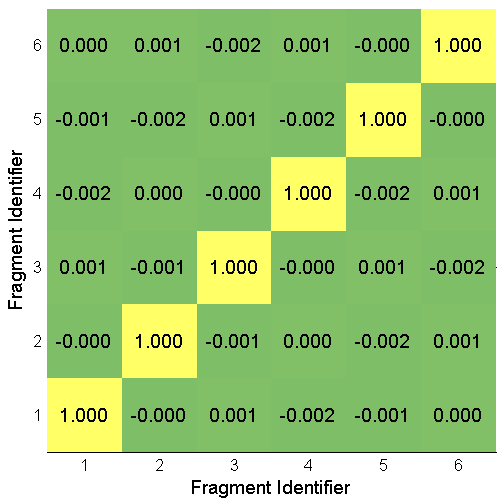}\label{fig:pdfk}} 
\caption{ {\it Correlation between fragments. Measured for $c=2$ (a) and $c=3$ (b). Block size was equal to 34 bytes. Correlation coefficients are close to 0, even between neighbor fragments.  }}%
\label{fig:correlation}%
\end{figure}

\noindent \textbf{Correlation.} Correlation coefficient was used to evaluate the linear dependence between data~\cite{correlation}. In Figure \ref{fig:correlation} correlation between fragments is shown. The method used for the calculation was same as in~\cite{coefficient:calcul}. Observed values of correlation coefficients are close to 0. It demonstrated that even neighbor fragments are not correlated with each other and thus confirmed the independence property of the scheme. 

\noindent \textbf{Difference.} Each fragment should be significantly different from the initial data and from other fragments of the same fragmentation result. Bit difference between a data sample and each of its fragments was measured and it was close to 50\%. Same result was obtained for the difference between fragments themselves.

\begin{table*}[h]
\caption{\textit{Runtime and storage requirements of different concerned fragmentation algorithm. $Poly\left(n,k,d\right)$: the cost of encoding data $d$ into $n$ fragments using a polynomial of degree $k-1$. $Matrix\left(n,k,d\right)$: the cost of multiplying data $d$ by a dispersal matrix $matrix$ of dimension $n \times k$. $Encrypt\left(d\right)$: the cost of encoding data using symmetric encryption. $Hash$: the cost of hashing the data. $RS$: the cost of applying a Reed-Solomon error correction code. $EncodeData\left(d,c,k\right)$: the cost of encoding data into fragments in our proposal. ($d$ - initial data, $|d|$ - initial data size, $key$ - symmetric key, $|key|$ - symmetric key size, $|b|$ - block size in our proposal, $|matrix|$ - matrix in IDA, $k$ - required number of fragments, $n$ - total number of fragments)}}
\label{tab:perf}
\begin{tabular}{|l|l|l|l|l|}
\hline
    Scheme & Runtime    &  Runtime  & Storage             & Storage   \\
           &  Fragmentation  &  Redundancy    &  Without Redundancy &  With Redundancy  \\ [0.03cm] \hline
    SSS     &  Poly(n,k,d)   &   -            &   $k|d|$            &   $n|d|$   \\ [0.02cm]  \hline
    IDA	    &  Matrix(n,k,d) &   -            &   $|d| + |matrix|$       &    $\frac{n}{k}|d| + |matrix|$ \\  [0.02cm]  \hline
    SSMS    &  Encrypt(d) + Poly(n,k,key)   &  RS(n-k,k,d)      &   $|d| + k|key|$  & $n\left(\frac{|d|}{k}+|key|\right)$  \\ [0.02cm] \hline
    AONT-RS &  Encrypt(d) + Hash(d)       &  RS(n-k,k,d)      &    $|d|+|key|$  &  $\frac{n}{k}(d+|key|)$  \\ [0.02cm] \hline
    Our proposal &  EncodeData(d,c,k) & RS(n-k,k,d)  & $|d| + k|pa|$  & $\frac{n}{k}\left(|d| + k|b|\right)$   \\ [0.02cm]  \hline
\end{tabular} 
\end{table*}

\section{Complexity Analysis and Storage Requirements}
\label{sec:theoretical-performance}
 
Table~\ref{tab:perf} shows an overview of complexity considerations  and storage requirements of concerned fragmentation schemes and compares them with the properties of our proposal. Algorithms may be divided into two groups. The first group relies on symmetric encryption for data protection and combines it with a key hiding or dispersal method that prevents the key (and therefore the initial data) recovery until $k$ fragments have been gathered. This group includes all variations of the all-or-nothing-transform and Secret Sharing Made Short. Second group, that includes Shamir's Secret Sharing and Information Dispersal, is not applying any symmetric key encryption, but encodes data using a system of equations, which is incomplete (and therefore harder to solve) when less than $k-1$ fragments are present. Such schemes may be quite fast as the encoding is based on simple operations (additions and multiplications). However, their big problem is the lack of scalability when the required number of fragment $k$ is growing. It comes from the fact that a growing $k$ entails a growing polynomial degree (SSS) or a growing dimension of the dispersal matrix (IDA), and in consequence a linear $O\left(k\right)$ increase of operations to be performed. Our algorithm overcomes the scalability issue by introducing the $c$ parameter. Data are dispersed over $k$ fragments, but encoded using a polynomial of degree $c$. What differentiates it also from the second group, is the fact that it does not provide availability at the same time as data protection. If needed, additional data fragments are obtained using the same $RS$ method than in encryption-based algorithms. The following subsections give more details about the complexity and storage requirements of analyzed algorithms. A precise performance evaluation is hard because of the variety of implementations. 

\subsection{SSS and IDA}
SSS computes $n$ values of a polynomial ($Poly$) for protecting a data $d$ of size $|d|$.  Thus, its performance depends on the values of $k$, $n$, and $|d|$. Evaluating a polynomial is usually done using the Horner's scheme, which is a $O\left(k\right)$ operation. The cost of an IDA equals to the cost of multiplying data $d$ by a $ k \times n$ dispersal matrix ($Matrix$). In both cases, data are usually first divided into smaller chunks and processed in a chunk by chunk fashion. They strongly benefit when the chunk size is equal to one byte, as it allows an implementation in finite field arithmetic of the field $GF\left(256\right)$. 

\subsection{SSMS and AONT-RS}
Performance of AONT-RS depends on the chosen encryption ($Encrypt$) and hash algorithms ($Hash$), as well as on the data size. If wisely implemented, SSMS applies the same mechanisms than AONT-RS: symmetric encryption ($Encrypt$) and Reed-Solomon ($RS$) code for redundancy. Instead of hiding the key inside the hash of the whole data, SSMS disperses it within the fragments using Shamir's scheme ($Poly$) applied only on the key. In the case when SSMS is applied on data much larger than a symmetric key, the time taken by the key fragmentation is negligible. To produce additional fragments, both schemes use Reed-Solomon error correction codes ($RS$). Performance of $RS$ may be slightly improved using Cauchy Reed-Solomon coding \cite{bib:blomer}.

\subsection{Our algorithm}
 
The fragmentation procedure $EncodeData\left(c,k,d\right)$ is composed of several steps: generating and splitting permutations, data distribution, data encoding, and data permuting. The most consuming operation is the mini-blocks encoding. It takes $c-1$ additions and $c-1$ multiplications to encode a single mini-block into a mini-share, as the Horner's scheme for evaluating a polynomial is used. The procedure is repeated for all the mini-blocks inside the data, so at the end $\#d\#b\left(c-1\right)$ additions and same number of multiplications are needed to encode the whole data. Because a $GF\left(256\right)$ finite field is used (like for SSS and IDA), a lookup table is used to replace the multiplications and the additions are implemented as a xor. Permuting a mini-share inside a share may be implemented as a constant time operation. Data dispersal function \textsc{Form\-Fragments} is an $O\left(k\right)$ operation. Permutations generation may be implemented in various ways and depends on the size $|pa|$. Permutations splitting depends on the size of $|pa|$ and the value of $c$. Being very simple and applied only once, data dispersal and permutations generation and splitting have a negligible effect on the algorithm performance. Additional fragments (if needed) are generated inside an optional procedure $RS$, which is exactly the same as the one used for SSMS and AONT-RS. The procedure results in a $k|pa|$ storage overhead (when $RS$ is not applied). A larger data block increases this overhead, but at the same time improves data protection and performance, as it allows a  better parallelization of encoding. Once the permutations are recovered, the defragmentation procedure is fully parallelizable, as a share may be decoded without waiting for the previous block to be processed.

\section{Measured performance}
\label{sec:measured-performance}

We benchmarked the proposed algorithm against the state-of-the art fragmentation techniques presented in Section \ref{sec:relevant}. All schemes were implemented in JAVA using following resources:  JDK 1.8 on DELL Latitude E6540, X64-based PC running on Intel\textsuperscript{\textregistered} Core\textsuperscript{TM} i7-4800MQ CPU @ 2.70 GHz with 8 GB RAM, under Windows 7.  The standard $javax.crypto$ library was used to implement cryptographic mechanisms. For each algorithm, the throughput was measured in an identical way on random data samples of 100MB.

\subsection{Implementation details}

Similarly to Shamir's scheme, the proposed algorithm can be implemented in any Galois Field $GF\left(2^Q\right)$. $Q$ is usually selected according to the word size of processors and can be 8, 16, 32 or 64-bit. The presented version was implemented in $GF\left(2^8\right)$, which allows the use of only logical operations. Same field was used for the implementations of IDA and SSS. AONT-RS was implemented in two versions: fast and secure. The fast version uses a combination of RC4 and MD5, the secure version is based on AES and SHA-256. Similarly, SSMS was implemented in two versions: one using RC4 and second using AES, and SSS was applied for key protection. For larger data the cost of splitting the key was negligible. We are aware that RC4 is an obsolete way of protecting data. However, it was used in the implementation of the original fast AONT-RS \cite{Resch:2011} and makes a good reference point for the speed comparison. The AES was used in the CTR mode with a 128 bits key and the AES-NI instruction set was enabled during the run.

\subsection{Results}

\begin{figure}[h]
\captionsetup{belowskip=-5pt}
\centering
\includegraphics[width=0.95\linewidth]{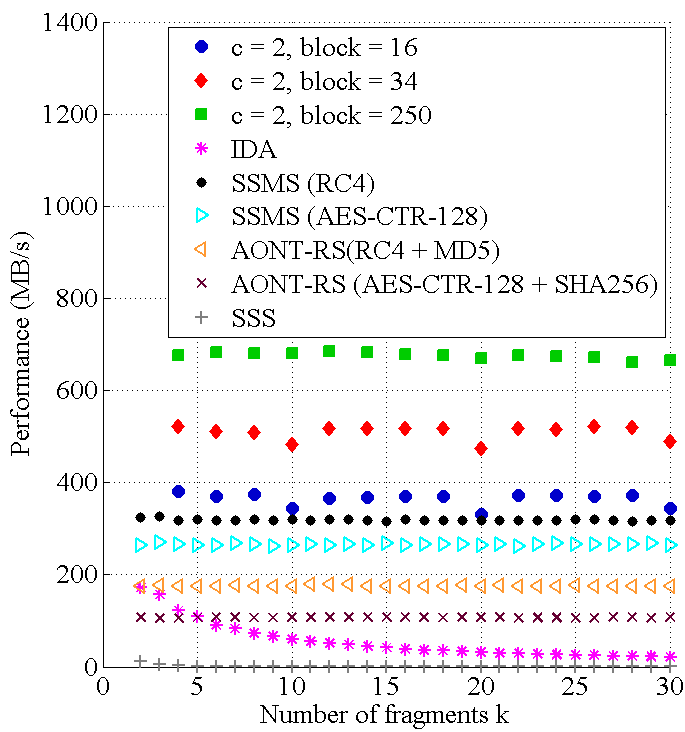}  
\caption{{\it Performance benchmark, $c = 2$. }}%
\label{fig:perfC2}%
\end{figure}

\begin{figure}[h]
\captionsetup{belowskip=-5pt}
\centering
\includegraphics[width=0.95\linewidth]{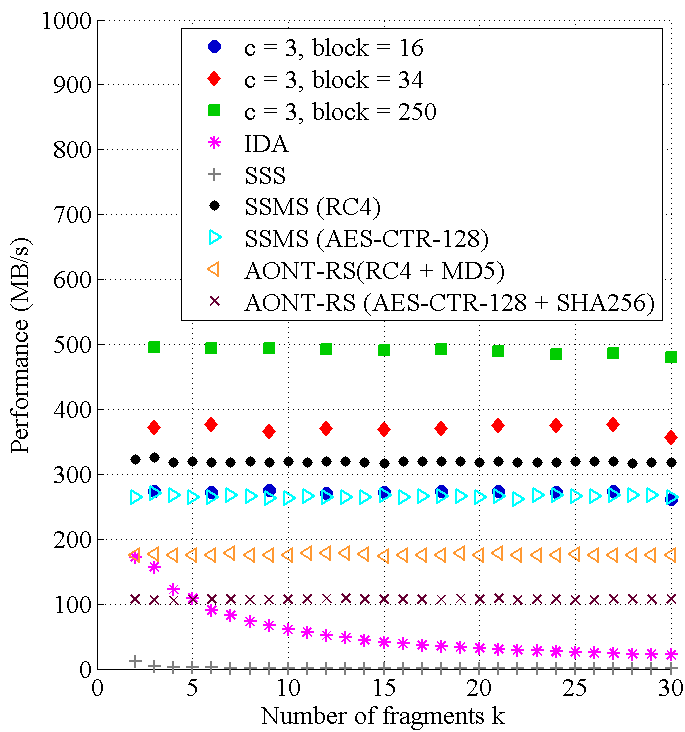}  
\caption{{\it Performance benchmark, $c = 3$. }}%
\label{fig:perfC3}%
\end{figure}

The fragmentation throughput was measured for six different configurations of our scheme: for two different values of $c$ (2 and 3) and three different choices of the block's size (16, 34, and 250 bytes). (A block size of 16 bytes was chosen as it produces the same storage overhead than SSMS with a 128-bits key. A block size of 34 bytes makes the recovery of a permutation array similar to performing a brute-force attack on a 128-bits key. A block size of 250 optimizes the performance of the scheme.) Figure \ref{fig:perfC2} and \ref{fig:perfC3} show the results of the comparison with relevant works. Our scheme is up to 50\% ($c=2$) or 40\% ($c=3$) faster than the fastest (SSMS with RC4) of the relevant works. It is more than 60\% ($c=2$) and 40\%($c=3$) faster in comparison to SSMS with AES. The cost of AONT-RS is higher than the cost of SSMS (as hashing data is more costly than splitting the key). In contrary to rest of the algorithms, IDA and SSS do not scale with the number of fragments $k$. 

\begin{figure}[!h]
\captionsetup{belowskip=-5pt}
\centering
\subfloat[][]{\includegraphics[width=0.5\linewidth]{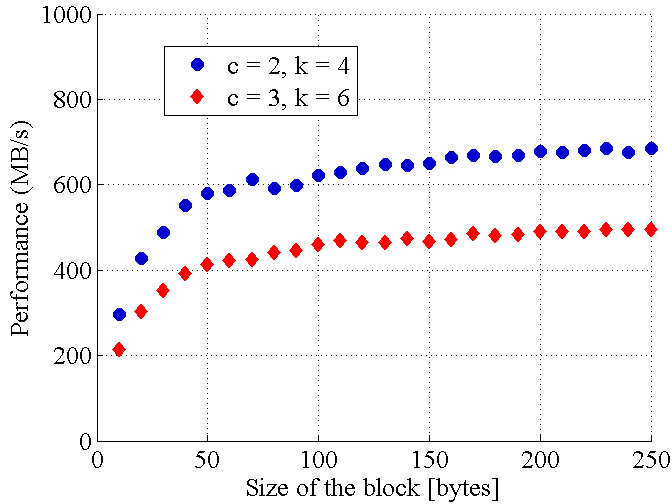} \label{fig:block}}
\subfloat[][]{\includegraphics[width=0.5\linewidth]{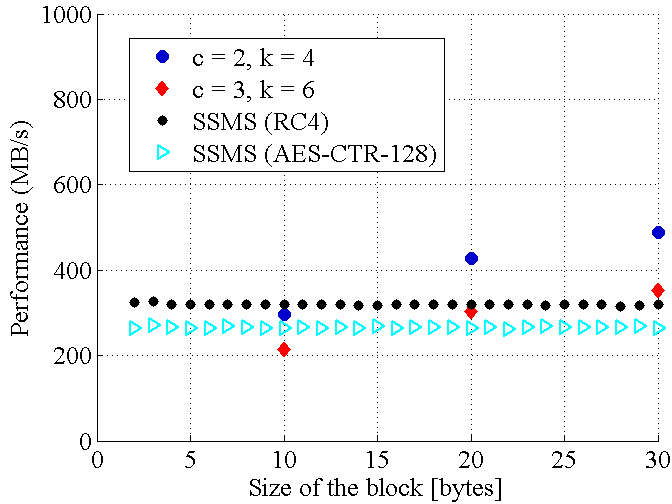}\label{fig:block-comparison}} 
\caption{ {\it Performance in function of the block size for $c=2$ and $c=3$ (a). A block size of 30 bytes allows the scheme to be faster than the relevant work (b) }}%
\label{fig:block}%
\end{figure}

As presented in Figure \ref{fig:block}, the chosen size of the block has an impact on the speed of our proposal. the fragmentation throughput grows with the the block size until achieving a maximum around 200-250 bytes. A block size of around 10 bytes ($c=2$) or 20 bytes ($c=3$) is enough to achieve the same performance as the two fastest of the relevant works (see Figure \ref{fig:block-comparison}). For a higher throughput of the fragmentation (and a better data protection level) a block size of more than 200 bytes is recommended.

\section{Future works}

In the future, we plan to perform a deeper cryptanalysis of our algorithm to precisely position its level of data protection. We also consider to implement a new version of the algorithm that would exploit the possibility of parallelization of the processing due to fragmentation . Moreover, the described fragmentation algorithm could be seen as a particular case of a more general method for data protection mixing fragmentation, data encoding, and data dispersal. Modifications in the way of data dispersal over fragment, data encoding, or data permuting could be done. For instance, data permutations can be done in a more complex and less regular way adding security (but probably loosing performance and capability of parallelization). As a second instance, in a pre-processing step data (i.e. like in \cite{Qiu:2015}) could be  first shred into two parts - a confidential and a non-confidential one - and then fragmented. It would generate fragments of different importance that could be dispersed in a selective manner over site with different levels of trustworthiness and cost. A different sharing scheme could be also applied for the data encoding. Last, but not least, the scheme could be adapted for different and more elaborated use cases, for instance involving several users. Indeed, fragmentation techniques are particularly well adapted to the nature of a multi-cloud environment, but they are also seen as an alternative to onion routing \cite{Katti:slicing} or as a key-less way of data protection in sensor networks. 

\section{Conclusion}

A novel algorithm for protecting data through fragmentation, encoding, and dispersal was introduced and analyzed. Data transformation into fragments relies on a combination of secret sharing and data permuting. During this process dependencies are created between data inside the fragments, in such a way that full data  recovery is possible only when all fragments have been gathered, which is possible only by acquiring the location and the different access rights of $c$ independent storage providers. First performance benchmarks show that the scheme is up to twice time faster than the state-of-the art comparable and widely renown techniques, while producing a reasonable storage overhead. Empirical security analysis verified that data inside the produced fragments do not preserve patterns and is therefore resistant to statistical analysis. The proposed scheme is particularly well adapted for data dispersal in a multi-cloud environment, where non-colluding cloud providers ensure the physical separation between the fragments. A method for an optimal secure dispersal of produced fragments between the clouds was presented. The scheme architecture is flexible, so it is possible to balance between desired performance, level of data protection, and storage overhead.


\bibliographystyle{ACM-Reference-Format}
\bibliography{samplebibliography} 

\end{document}